\documentclass[12pt]{article}
\usepackage{amsmath}
\usepackage{amssymb}
\usepackage{amsfonts}
\usepackage{epsfig}

\newcommand{\rhoext}{\rho_{\rm ext}}

\def\gsim{\mathrel{\raise.5ex\hbox{$>$}\mkern-14mu
             \lower0.6ex\hbox{$\sim$}}}

\def\lsim{\mathrel{\raise.3ex\hbox{$<$}\mkern-14mu
             \lower0.6ex\hbox{$\sim$}}}

\begin{document}

\title{Compton dragged supercritical piles: The GRB prompt and afterglow scenario}
\author{A. Mastichiadis$^1$, D. Kazanas$^{2}$ \\
\ \ \ \ \\
$^{1}$\textit{Department of Physics }\\
\textit{University of Athens,  }\\
\textit{Panepistimiopolis,   GR 15783, Zografos, Greece.} \ \\
\ \ \ \ \ \ \ \ \ \ \\
$^{2}$\textit{NASA/Goddard Space Flight Center,}\\
\textit{Mail Code 663, Greenbelt, MD 20785}}
\date{ }
\maketitle

\abstract{ We examine the prompt and afterglow emission
within the context of the Supercritical Pile model for GRBs.
For this we have performed self-consistent calculations
by solving three time-dependent kinetic equations for protons,
electrons and photons in addition to the usual mass
and energy conservation equations. We follow the evolution
of the RBW as it sweeps up circumstellar matter and assume
that the swept-up electrons and protons have
energies equal to the Lorentz factor of the flow.
While the electrons radiate their energies through
synchrotron and inverse Compton radiation on short
timescales, the protons, at least initially,
start accumulating without any dissipation. As the 
accumulated mass of relativistic protons
increases, however, they can become supercritical
to the `proton-photon pair-production - synchrotron radiation'
network, and, as a consequence, they transfer explosively their
stored energy to  secondary electron-positron pairs and radiation.
This results in a burst which has many features similar
to the ones observed in GRB prompt emission.
We have included in our calculations the radiation drag
force exerted on the flow from the scattered radiation
of the prompt emission on the circumstellar material.
We find that this can decelerate the flow on
timescales which are much faster than the ones related to
the usual adiabatic/radiative ones. As a result the emission
exhibits a steep drop just after the prompt phase,
in agreement with the Swift afterglow observations.}



\section{Introduction}

Despite the great progress in the GRB field made with {\sl CGRO} and 
{\sl BeppoSAX} (\cite{Costa97}),
there are  major issues concerning the dynamics and radiative
processes of these events that still remain open.
Some of these have been with us since the inception of
the cosmological GRB models, while others are rather new, 
the outcome of the wealth of new observations made by 
{\sl Swift} and {\sl HETE}.
Chief among them
are the conversion of the RBW kinetic energy into radiation, the
fact that the photon energy at which the GRB luminosity peaks
is narrowly distributed around a value intriguingly
close to the electron rest mass energy 
and the transition from the prompt to the afterglow emission.
The purpose of the present
note is to describe a process that  provides a ``natural'' account
of these generic, puzzling GRB features.

\section{Blast Wave Dynamics and Radiation}

We consider a Relativistic Blast Wave (RBW)
moving with speed $\upsilon_{0}=\beta_\Gamma c$,
where $\beta_{\Gamma} =(1-\Gamma^{-2})^{-1/2}$ and $\Gamma$ the
bulk Lorentz factor of the flow. It has a radius $R(t)$
as measured from the origin of our coordinate system (assumed to
be the center of the original explosion) and it is sweeping
mass of density $\rhoext$. As the RBW sweeps up mass
from the circumstellar matter (CSM), it starts slowing down.
Following \cite{CD99} we write
two equations for the evolution of the RBW. One for the mass
\begin{equation}\label{equ1}
{dM\over{dR}}=4\pi R^2\Gamma\rhoext - {1\over{c^3\Gamma}}\dot E
\end{equation}
and one for the Lorentz factor $\Gamma$, which reflects the 
energy-momentum conservation
\begin{equation}\label{equ2}
{d\Gamma\over{dR}}=-{4\pi R^2\rhoext\Gamma^2\over M}-{F_{rad}\over{Mc^2}}.
\end{equation}
Here $\dot E$ is the radiation rate as measured in the comoving frame
and $F_{rad}$ is the radiation drag force which is exerted on the RBW
from any radiation field exterior to the flow.
In the standard case (for a review see \cite{Pir05})
the above equations specify completely the dynamics
of the RBW once the initial conditions of the flow
have been specified.
The profile of the external density $\rhoext$ has also to be specified
-- in most cases either a constant density or a wind profile is assumed.
The radiation rate $\dot E$ is usually set $\sl a-priori$
varying between two extremes: When there is no radiation,
the flow is considered as adiabatic, while when the hot mass is
immediately radiated away the flow is considered as radiative.
Either choice has a profound influence on the evolution of the RBW:
In the adiabatic case the solution of Eqns (1) and (2) gives $\Gamma\propto R^{-3/2}$
while in the radiative case one gets $\Gamma\propto R^{-3}$.
On the other hand, the role of $F_{rad}$ has not so far been investigated.
 Assuming that this force is exerted on
the flow by the RBW photons scattered on the CSM,
we can write (Mastichiadis \& Kazanas, in preparation)
\begin{equation}\label{equ3}
F_{rad}={{64\pi}\over{9c}}\tau_b n_e^{CSM}\sigma_T R\Gamma^4\dot{E}
\end{equation}
where $\tau_b$ is the Thomson optical depth of the RBW,
$n_e^{CSM}$ is the electron density of the CSM and $\sigma_T$
is the Thomson cross section.

In order to calculate self-consistently
the $\dot E$ and $F_{rad}$ terms,
we implement a numerical code for the radiation transfer
in the RBW. Details have been given in \cite{MK95} and \cite{MK06}
but for the sake of completeness we repeat here the
basic points about it.

The equations to be solved can be written in the general form
\begin{equation}\label{equ4}
{{\partial n_i}\over{\partial t}} + L_i +Q_i=0.
\end{equation}
The unknown functions $n_i$ are the differential  
number densities of protons, electrons and photons
while the index $i$ can be any one of the subscripts `p', `e'
or `$\gamma$' referring to each species.
The operators $L_i$  denote losses and escape from the system
while $Q_i$ denote injection and source terms. The important
physical processes to be included in the kinetic equations are
proton-photon (Bethe-Heitler) pair production,
photopion production, lepton synchrotron radiation, synchrotron self absorption, 
inverse Compton scattering (in both the Thomson and Klein-Nishina regimes)
and photon-photon pair production.
The processes involving photons
can take place either with
photons produced directly or with the photons
which have been reflected by the CSM
in front of the advancing shock.
The above equations are solved in the fluid frame inside a spherical blob of
radius $R_b=R/\Gamma$. This can be justified
from the fact that due
to relativistic beaming an observer receives the radiation coming mainly
from a small section of the RBW of lateral width $R/\Gamma$ and
longitudinal width $R/\Gamma^2$ in the lab frame but $R/\Gamma$
on the comoving frame.

\begin{figure}
\begin{center}
\includegraphics [width=0.6\textwidth]{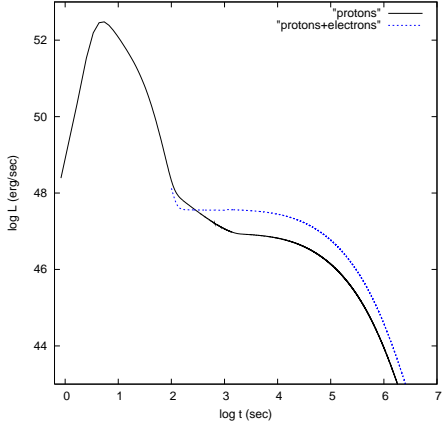}
\end{center}
\caption{Photon luminosity as a result of a proton supercriticality
which is developed as the RBW sweeps up mass from the CSM. 
For the initial values of the run see text.}
\end{figure}

Despite the fact hat the general frame adopted here is similar
to \cite{MK06}, the present approach differs in
two important aspects:

(1) Hot protons accumulate continuously on the RBW. For this
we introduce injection terms for protons  and
electrons, assuming that at each radius $R$ the RBW picks
up an equal amount of electrons and protons from the
CSM  which have, upon injection,  energies  
$E_p=\Gamma m_pc^2$ and $E_e=\Gamma m_ec^2$ respectively.   
Consequently, the proton {\sl energy} injection rate
is given by \cite{BMK76}
\begin{equation}\label{equ7}
\left({{dE}\over {dt}}\right)_{inj}=4\pi R^2\rhoext(\Gamma^2-\Gamma)c^3
\end{equation}
while a fraction $m_e/m_p$ of it goes to electrons.

(2) The scattering of the RBW photons takes place
on the CSM matter in front of the advancing front,
thus the column density is also uniquely determined
from the initial conditions. In this way we relax
the assumption of \cite{MK06} about the optical depth
of the mirror.

Eqns (1), (2) and (4), with the addition of Eqns (3) and (5) 
form a set of equations which can now be solved
simultaneously to give us the evolution of the RBW
dynamics and radiated power. We should emphasize that this
approach is self-consistent as the 'hot' mass injected
through Eqn (5) corresponds to
the RHS of Eqn (1) while the radiated luminosity $\dot E$
and the radiative force $F_{rad}$
are calculated from Eqns (4). Therefore, once the initial
conditions which are (i) the energy of the explosion $\cal E$,
(ii) the initial radius at which the RBW starts sweeping up
matter $R_0$, (iii) the value  $\Gamma_0=\Gamma(R_0)$
and (iv) the external density of the CSM are specified,
one can solve the above set of equations forward in time.
However, the radiation code
needs one more parameter, which is the value
of the magnetic field
at position $R_0$ and an assumption on its profile with $R$.
Without loss of generality we can assume
that the magnetic field drops like $1/R$
as assumed in \cite{VK03}.

\begin{figure}
\begin{center}
\includegraphics [width=0.9\textwidth]  {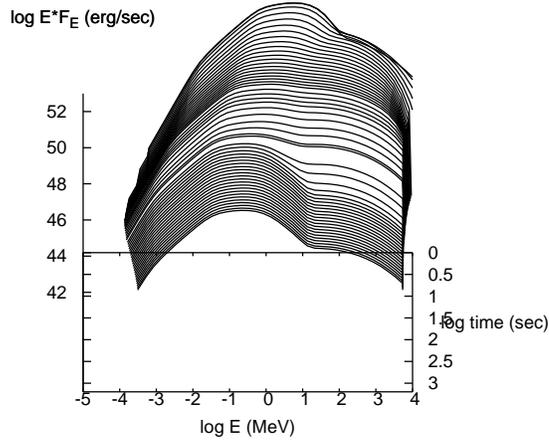}
\end{center}
\caption{Evolution of the bulk Comptonized spectrum during the outburst shown in Fig.1.}
\end{figure}

\section{Proton Supercriticality}

As emphasized in \cite{KM92} and \cite{KGM02}
relativistic protons can become supercritical in a
network involving proton-photon
(Bethe-Heitler) pair production and electron synchrotron    
radiation. This supercriticality is a
radiative-type
instability which can convert the free energy of the relativistic proton
plasma into relativistic $e^+ e^-$ pairs, once 
a kinematical and a dynamical criterion are simultaneously
satisfied.

For the kinematical criterion to be fulfilled 
all one needs is that the synchrotron photons produced
from the Bethe-Heitler pairs  to be
energetic enough as to produce more pairs on the
relativistic protons (see \cite{KM92}). As it was shown in \cite{KGM02},
if the proton plasma itself
is in relativistic bulk motion this criterion is satisfied if
\begin{equation}\label{equ6}
b\Gamma^5\ge 2
\end{equation}
where $b=B/B_{crit}$ with $B_{crit}
= m_e^2 \, c^3/\hbar c$ = $4.4 \; 10^{13}$ G,
the critical magnetic field.

In our case, if the choice of the values of
these parameters are such that the above criterion is
not fulfilled initially, then it will not be
fulfilled anywhere, provided that $\Gamma$ is
in its coasting stage -- as implied by eqn.(2).
This happens because, as  $B$ and $\Gamma$
drop outwards, $b\Gamma^5$ can only decrease as $R$ increases.
Therefore this corresponds to the trivial case:   
The protons are accumulated at the rate given by
eqn (1) and they remain inert no matter how much mass
has been accumulated on the RBW.
The electrons, on the other hand,
can radiate their energies fast as their radiative lifetimes
to synchrotron and/or ICS
can be short compared to the light crossing timescale on the RBW frame.
However, the produced  $\dot {E}$
is rather low and cannot affect significantly the global
eqns (1) and (2).
Thus the RBW obeys, for all practical purposes, the non-radiative
case, while its luminosity stays at a relatively low level.

Far more interesting is the case where the kinematical
criterion is satisfied initially. Then whether the
flow becomes radiatively unstable will depend on the second
criterion. Qualitatively one can say that this 
criterion is satisfied if 
{\sl at least one} of the
synchrotron photons produced by the $e^+ \, e^-$ pairs produces
another pair before escaping the volume of the plasma in a reaction
with a sufficiently energetic proton (i.e. one that fulfills the
kinematic threshold). This results in a condition for the column   
density of the plasma which is identical to that of a critical
nuclear chain reaction
and can  be examined  numerically  
with the code described in the previous section.
Running various cases
we can find those initial conditions which can make the
flow unstable.

Figure 1 shows the photon evolution in the lab frame
of a solution that becomes supercritical. 
The initial values of the run are 
$R_0=10^{15}$cm, 
$n_{ext}=10^5$ par/cm$^3$ 
(values consistent with those of the wind of
a WR-star),
${\cal E}=10^{54}$ erg, 
$\Gamma_0=200$,
and $B_0=10$ G. Thus at the beginning of the
swept-up phase
the product $b_0\Gamma_0^5\simeq 7$ and the kinematic criterion (6) is
satisfied. The swept-up protons become quickly supercritical 
and the photons grow exponentially. This eventually leads to fast proton
cooling and photon saturation leading to the peak in the
photon flux of Fig.1.
Consequently an abrupt decrease in $\Gamma$
due to the $F_{rad}$ term in Eqn (2) leads to a steep decay of photons
for the next hundreds of seconds. As the photon density decreases,
lower energy protons continue cooling in a more controlled manner.
This gives a characteristic flattening to the photon lightcurve 
for the next few thousands seconds. At even longer timescales 
(tens to hundred thousand of seconds) first the strength of the
magnetic field $B$ and later $\Gamma$ start decreasing due to 
the RBW expansion. These effects cause a final drop in the observed
photon luminosity. Thus, this model can, in principle, connect
the prompt to the afterglow emission. 

As it was stated in \cite{KGM02} and shown numerically
in \cite{MK06}
the observed photon spectrum
consists of two components: one that is observed directly
and one that is produced from the bulk Comptonization
of the aforementioned direct component scattered first on the CSM
electrons and then on the cool pairs of
the RBW. For loops operating close to threshold, this has 
a peak at $\simeq$ 1 MeV independent of the initial value of $\Gamma$.

Figure 2 shows snapshots of the bulk Comptonized spectrum
for the first thousand seconds after the photon burst
for the run described above.
This has a peak in the MeV region at maximum luminosity.
Consequent proton cooling and $\Gamma$ reduction 
have as effects (i) the fast decrease of the luminosity,
as already shown in Fig.1 and (ii) a softening of the
spectrum from the MeV to sub-MeV / keV regime.

\section{Acknowledgements}
The project is 
co-funded by the European Social Fund 
and National Resources - (EPEAEK II) PYTHAGORAS 
and an INTEGRAL Theory Grant.

\end{document}